\documentclass[submission,copyright]{eptcs}
\providecommand{\event}{INFINITY 2012} 
\usepackage{breakurl}                  

\usepackage[T1]{fontenc}
\usepackage{ae}

\usepackage{bold-extra}
\usepackage{tikz}
\usetikzlibrary{shadows,automata,petri,shapes}
\usepackage{xcolor}
\usepackage{stmaryrd}

\usepackage{amsmath}
\usepackage{amssymb}

\usepackage{url}
\usepackage{subfigure}
\usepackage{enumerate}

\usepackage{algpseudocode}
\usepackage{algorithm}

\newtheorem{definition}{Definition}[section]
\newtheorem{theorem}{Theorem}[section]
\newtheorem{proposition}{Proposition}[section]

\def\qed{\ifmmode\squareforqed\else{\unskip\nobreak\hfil
  \penalty50\hskip1em\null\nobreak\hfil$\Box$%
  \parfillskip=0pt\finalhyphendemerits=0\endgraf}\fi}

\def\st{~|~} 
\def\defeq{\mathrel{\smash{\stackrel{{\scriptscriptstyle\mathrm{df}}}{=}}}}
\def\set #1{\ensuremath{\mathbb{\uppercase{#1}}}}
\def\tuple#1{\ensuremath{
    \langle #1 
    \rangle }%
}
\def\dom{\mathit{dom}}
\def\childarrow#1{\mathop{\smash{\xrightarrow{~#1~}}}}

\def\path{\mathit{path}}
\def\paths{\mathit{paths}}
\def\relpath#1{\widetilde{\mathit{path}}(#1)}

\def\trees{\ensuremath{Trees}}

\def\seq#1#2#3{{#1}_{#2}, \dots, {#1}_{#3}}
\def\parent{\ensuremath{\sphericalangle}}
\def\sibling{\ensuremath{\mathbin{{\pitchfork}}}}
\def\markings{\ensuremath{\mathit{Mrk}}}
\def\generatorplace{s_{\eta}}
\def\Places{\ensuremath{\mathit{S}}}
\def\pidtreeset{\Xi}

\def\siblingequiv{\sim}
\def\siblingequivh#1{\sim_{#1}}

\def\ie{\emph{i.e.}}

\def\prefix{\mathit{prefix}}
\def\subpid{\mathit{subpid}}
\def\next{\mathit{next}}
\def\nextpid{\mathit{nextpid}}
\def\pid{\mathit{pid}}

\def\fire #1{{[#1\rangle}}

\title{Effective Marking Equivalence Checking in Systems with Dynamic Process Creation\footnote{This work has been granted by project Synbiotic ANR BLANC 0307 01.}}
\def\titlerunning{Effective Marking Equivalence Checking}

\author{\L ukasz Fronc
  \institute{IBISC, Université d'Evry-Val d'Essonne\\
    23 boulevard de France, \\
    91037 Évry, France}
  \email{lfronc@ibisc.univ-evry.fr}
}
\def\authorrunning{\L ukasz Fronc}

\begin{document}

\maketitle

\begin{abstract}
  The starting point of this work is a framework allowing to model
  systems with dynamic process creation, equipped with a procedure to
  detect symmetric executions (\ie, which differ only by the
  identities of processes). This allows to reduce the state space,
  potentially to an exponentially smaller size, and, because process
  identifiers are never reused, this also allows to reduce to finite
  size some infinite state spaces.
  However, in this approach, the procedure to detect symmetries does
  not allow for computationally efficient algorithms, mainly because
  each newly computed state has to be compared with every already
  reached state.

  In this paper, we propose a new approach to detect symmetries in
  this framework that will solve this problem, thus enabling for
  efficient algorithms. We formalise a canonical representation of
  states and identify a sufficient condition on the analysed model
  that guarantees that every symmetry can be detected. For the models
  that do not fall into this category, our approach is still correct
  but does not guarantee a maximal reduction of state space.
\end{abstract}

\section{Introduction}

The problem of detecting symmetries during the construction of state
spaces has been widely explored \cite{Junttila}. Given a formalism
and its operational semantics, the state space is built starting from
an initial state. We repeat the computation of new states for every
discovered state and we aggregate them in order to reach a fixed
point, which is the set of all reachable states. To add abstraction to
our formalism we can define equivalence classes of states, two states
in the same class are said to be symmetric. Each time a state
symmetric to an already visited one is found, it is considered as
already known, also it means that if the abstraction was well defined,
then the behavior resulting from further exploration of this state
leads to an already analysed behaviour and thus may be omitted. The
resulting state space corresponds to the quotient graph of equivalence
classes of states (symmetric states) and can be exponentially smaller
than the original graph \cite{Clarke98}, or more drastically infinite
state spaces can be reduced to finite ones if the number of
equivalence classes is finite. Reductions by symmetries have been
implemented in a variety of tools and proved to be a successful
technique \cite{Hendriks04addingsymmetry,Kenneth1993,BosnackiDH02}.

Here we focus on symmetries induced by dynamic process creation in
multi-threaded computing systems. The goal of formalisms using this
paradigm is to allow processes to be created at runtime. This ability
contributes to the size explosion of state spaces, in particular the
order of created processes, despite being irrelevant, introduces
large amounts of interleaving. Moreover processes may be created
endlessly which would lead to infinite state spaces if states cannot
be identified as equivalent. In the formalism we focus on, process
names are irrelevant, it is the relations between them that matter
and not their identities. For example, let us consider an
\texttt{http} server, each time it receives a request for the home
page of a website, it creates a new thread which will handle the
request. Now suppose that the request has been handled but the server
receives another one for the same home page, then it creates a
\emph{new} thread which will handle the new request. These two threads
have different identities however they answer the same kind of requests
in same conditions, the resulting behaviour will be the same modulo
thread identifiers, they are clearly symmetric. This example well
illustrates the nature of symmetries we want to detect.\looseness=-1

The current work is based on the approach developed in
\cite{KKPP08,KKPP09,KKPP10} which presents a method to detect
symmetries introduced by process identifiers. The approach in these
works reduces the detection of symmetries to computation of graph
isomorphisms. Each newly discovered state is transformed into a graph,
if this graph is isomorphic to any graph corresponding to a visited
state then the two states are symmetric. The main problem with this
approach is that each time we discover a state, we have to look for an
isomorphism with all the visited states.

In this paper we present an effective way to detect symmetries without
computing a graph isomorphism. It provides representations for
markings that allow direct comparison of states, with a hash table for
example, instead of requiring a comparison with all visited
states. Let us consider the following state space exploration
algorithm, similar algorithms can be found in \cite{FP11a,FP11}. We
compute the state space of a model starting with an initial state in a
set $\mathit{todo}$. For each state $s$ in set $\mathit{todo}$, we
check if the state has already been visited using a function
$\textsc{visited}$. If the state was not visited yet then we add it to
a set $\mathit{done}$ and add all of its successor states (by calling
for example a function $\textsc{succs}(s)$) to set $\mathit{todo}$. At
the end of execution the set of reachable states is the set
$\mathit{done}$. Symmetry detection can be performed in function
$\textsc{visited}$. If we use the graph isomorphism approach, we have
to check for an isomorphism with all states in set $\mathit{done}$,
whereas using canonical representation we will just test if the state
\emph{is} in set $\mathit{done}$. The former approach is more
expensive due to the involved loop and the isomorphism computation
even if this operation may be fast with distinct graphs
\cite{nauty1,nauty2}. With the latter we can perform the test in
constant time using a hash table for example, indeed the
representation being canonical we can build a hash function for table
lookup and then use a few comparisons.

The method presented here does not detect all symmetries whereas the
graph isomorphism method does. However, we trust that we detect a
large amount of symmetries in systems where completeness is not
achieved and the trade-off between completeness and efficiency is
interesting. Moreover, we provide a sufficient condition for maximal
reductions, \ie, for completeness.

This paper is structured as follows: first we introduce process
identifiers and recall the definition of the Petri net models we use,
next we introduce the data structure we will use to represent
markings, then we present the theoretical basis for symmetry detection
with our representation, and finally we discuss canonisation of the
representation and a sufficient condition to achieve completeness. Due
to space limitation, proofs have been omitted but a version of this
paper including the proofs is available as a technical report~\cite{fullpaper}.

\section{Process identifiers and Petri nets}

\subsection{Process identifiers}

Addressing systems with multiple threads or multiple processes is a
tedious task. When the number of processes is fixed, each process
can be represented as a subnet in the model. This approach is limited,
indeed the number of processes is fixed and thus no new processes can
be created. In real life applications we create processes at runtime
and so we need to reason about this dynamic process creation and the
resulting behaviour.

The problem of reasoning about this kind of systems has lead to the
development of different formalisms and methods \cite{enlighten3197}.
Here, we focus on an implementation in terms of Petri nets
\cite{Jensen09} which allows for dynamic process creation and is
mainly based on \cite{KKPP10}. First we refine the definition of
process identifiers that will be used through this paper, the
definition is compatible with the one from \cite{KKPP10}. The only
difference is the introduction of an empty pid that helps the formal
definition of pids. Next, we also add some useful operations to handle
pids.

\begin{definition}[process identifiers] Process identifiers (pids) are
  elements of $\set{P} \defeq (\set{N^+})^{\star}$, the set of tuples
  over non-zero natural numbers. $\set{P}$ equipped with the tuples
  concatenation operation and its identity element, the empty tuple
  \tuple{}, is a monoid.  We denote $a_1 . a_2 . \, \cdots \, . a_n$
  the process identifier $\tuple{a_1, a_2, \dots, a_n}$.
\end{definition}
\def\length{\mathit{length}}

\begin{definition} Let $\pi$ be a pid. We define the length, the
  prefix and the set of subpids of $\pi$ as:
  \begin{itemize}
  \item $\length(\pi) = n$ if $\exists n > 0$ such that $\pi =
    \tuple{\seq{a}{1}{n}}$ and $0\!$ otherwise;
  \item $\prefix(\pi) = \tuple{\seq{a}{1}{n-1}}$ if $\exists n > 0$
    such that $\pi = \tuple{\seq{a}{1}{n}}$ and $\tuple{}$ otherwise;
  \item $\subpid(\pi) = \{ \pi \} \cup \subpid(\prefix(\pi))$ if
    $\length(\pi) > 0$ and $\emptyset$ otherwise.
  \end{itemize}
\end{definition}

Now as in \cite{KKPP10} we define the operations that can be used to
compare process identifiers in a model. The following operations are
the \emph{only ones} that can be used in the model to compare pids,
any other operation is forbidden.

\begin{definition}[operations on process identifiers] Two pids $\pi,
  \pi' \in \set{P}$ can be compared using the equality and the
  following operations:
  \begin{itemize}
  \item
    \begin{tabular}{p{0.1em}p{0.4em}p{1em}p{1em}p{7.5em}p{4em}l}
      $\pi$ & $\parent_1$ & $\pi'$ & iff & $\exists a \in \set{N}^+$ &
      such that & $\pi.a = \pi'$;
    \end{tabular}
  \item
    \begin{tabular}{p{0.1em}p{0.4em}p{1em}p{1em}p{7.5em}p{4em}l}
      $\pi$ & $\hspace{0.1em}\parent$ & $\pi'$ & iff & $\exists
      \seq{a}{1}{n} \in \set{N}^+$ & such that & $\pi.a_1. \cdots .a_n
      = \pi'$;
    \end{tabular}
  \item
    \begin{tabular}{p{0.1em}p{0.4em}p{1em}p{1em}p{7.5em}p{4em}l}
      $\pi$ & $\sibling_1$ & $\pi'$ & iff & $\exists \pi'' \in
      \set{P}, \exists i \in \set{N}^+$ & such that & $\pi'' \neq
      \tuple{},~ \pi = \pi''.i
      ~~\mbox{and}~~ \pi' = \pi''.(i+1)$;
    \end{tabular}
  \item
    \begin{tabular}{p{0.1em}p{0.4em}p{1em}p{1em}p{7.5em}p{4em}l}
      $\pi$ & $\hspace{0.1em}\sibling$ & $\pi'$ & iff & $\exists \pi''
      \in \set{P}, \exists i, j \in \set{N}^+$ & such that & $\pi'' \neq
      \tuple{},~ \pi =
      \pi''.i, ~ \pi' = \pi''.j ~\mbox{and}~ i < j$.
    \end{tabular}
  \end{itemize}
\end{definition}

Intuitively, $\pi\,\parent_1 \,\pi'$ means that $\pi$ is the parent of
$\pi'$, $\pi\,\parent\,\pi'$ means that $\pi$ is an ancestor of
$\pi'$, $\pi \sibling \pi'$ means that $\pi$ is a younger sibling of
$\pi'$, \ie, was spawned before $\pi$ and have the same parent as
$\pi$, finally $\pi \sibling_1 \pi'$ means that $\pi$ is the younger
sibling of $\pi'$ spawned just before $\pi'$.

The last refinement on pids is the introduction of a total order
between pids. This order will be important when addressing the
detection of symmetries and canonical forms.

\begin{definition}[ordering]
  The set of all pids $\set{P}$ is totally ordered with a hierarchical
  order, \ie, pids are ordered by length and if they have the same
  length, the lexicographic order on tuples is used.
\end{definition}

\subsection{Coloured Petri nets}

The models we address are (coloured) Petri nets which allow dynamic
process creation. A formal definition was given in
\cite{KKPP08,KKPP09} then refined in \cite{KKPP10}, we base ourselves
on the refined one. Here, we recall the definition and some
requirements. Let $\set{V}$ be a set of variables, $\set{D}$ a set of
data values and $\set{E}$ a set of expressions such that $\set{V} \cup
\set{D} \subseteq \set{E}$. We assume that $\set{E}$ contains Boolean
expressions. A binding is a partial function $\beta : \set{V}
\rightarrow \set{P} \cup \set{D}$. The application of a binding
$\beta$ is extended to denote $\beta(e)$, the evaluation of an
expression $e$ under $\beta$. The evaluation of an expression under a
binding is naturally extended to sets and multisets of
expressions. Data values, variables, syntax for expressions, possibly
typing rules etc. denote the colour domain of a Petri net.

\begin{definition}[Petri nets] A Petri net is a tuple $(S, T, \ell)$
  where $S$ is a finite set of places, $T\!\!$, disjoint from $S$, is a
  finite set of transitions, and $\ell$ is a labelling function such
  that:
\begin{itemize}
\item for all $s \in S$, $\ell(s)$ is the type of $s$, a Cartesian
  product $X_1 \times \dots \times X_k$ ($k \ge 1$), where each $X_i$
  is $\set{P}$ or $\set{D}$, this type denotes the values that $s$ may
  contain;
\item for all $t \in T$, $\ell(t) \in \set{E}$ is the guard of t,
  i.e., a condition for its execution;
\item for all $(x, y) \in (S \times T) \cup (T \times S)$, $\ell(x,
  y)$ is a multiset over $\set{E}$ and defines the arc from $x$ to $y$.
\end{itemize}
A \emph{marking} of a Petri net is a map that associates to each place $s \in
\Places$ a multiset of values from $\ell(s)$. We denote by $\markings$
the set of all markings. From a marking $M$, a transition $t$ can be
fired using a binding $\beta$ and yield a new marking $M'$, which
is denoted by $M\fire{t, \beta}M'$, iff:
\begin{itemize}
  \item there are enough tokens: for all $s \in \Places$, $M(s) \ge
    \beta(\ell(s, t))$;
  \item the guard is satisfied: $\beta(\ell(t))$ is true;
  \item place types are respected: for all $s \in \Places,~
    \beta(\ell(t, s))$ is a multiset over $\ell(s)$;
  \item $M'$ is $M$ with tokens consumed and produced according to the
    arcs: for all $s \in \Places, M'(s) = M(s) - \beta(\ell(s, t)) +
    \beta(\ell(t, s))$.
\end{itemize}
\end{definition}

\newtheorem{assumption}{Assumption}

The following requirements are adapted from \cite{KKPP10} which
generalised \cite{KKPP08,KKPP09} in two ways: first by using
transitions systems in general instead of Petri nets in particular;
second, by relaxing the constraints. We have preferred to stay with
Petri nets to extend our previous works~\cite{FP11,FP11a}. A Petri $N$
net respecting all the following requirements is called a \emph{thread
  Petri net} (or \emph{t-net}).
\begin{enumerate}
  \item The set of places of $N$ contains a unique generator place
    $\generatorplace$ having type $\set{P} \times \set{N}$. The
    generator place stores tokens $\tuple{\pi, i}$ where $i$ is the
    counter of child threads already spawned by $\pi$. Thus the next
    threads created by $\pi$ will have pids $\pi.(i+1)$, $\pi.(i+2)$,
    etc. We say that $\pi$ is \emph{generative} at a marking $M$ if there
    is a $n \in \set{N}$ such that $\tuple{\pi, n} \in
    M(\generatorplace)$.
  \item We assume that the initial marking $M_0$ of $N$ is such that
    the generator place contains exactly one token, $\tuple{\tuple{1},
      0}$, and all the other places are empty or contain data values.
  \item For each transition $t \in T$, the annotation on the arc from
    the generator place to $t$ is a set of the form
    $\ell(\generatorplace, t) \defeq \{ \tuple{p_1, c_1}, \dots,
    \tuple{p_k, c_k}\}$ where $k \ge 0$ and all $p_i$'s and $c_i$'s
    are distinct pid and counter variables. The annotation on the arc
    from $t$ to the generator place is a set: 
    $$
    \ell(t, \generatorplace) \defeq \left\{
    \begin{array}{l}
      \tuple{p_1, c_1+n_1}, \dots, \tuple{p_m, c_m+n_m}, \\
      \tuple{p_1.(c_1+1), 0}, \dots, \tuple{p_1.(c_1+n_1), 0}, \\
      \dots \\
      \tuple{p_k.(c_k+1), 0}, \dots, \tuple{p_k.(c_k+n_k), 0}
    \end{array}
    \right\}
    $$
    where $m \le k$, and $n_j \ge 0$ for all $j$. An empty arc
    annotation means that the arc is absent.  Below we denote by
    $\Pi_t$ the set of all newly created pids $p_i.(c_i + j)$ used in
    $\ell(t, \generatorplace)$.
  \item For each transition $t \in T$ and each place $s \in S
    \setminus\!\{ \generatorplace \}$, the annotation on the arc from
    $s$ to $t$ is a multiset of vectors built from variables and data
    values, and the annotation on the arc from $t$ to $s$ is a
    multiset of vectors built from expressions involving data
    variables and data values as well as elements from $\Pi_t \cup \{
    \seq{p}{1}{m} \}$.
  \item For each transition $t \in T$, $\ell(t)$ is a computable
    Boolean expression, built from the variables occurring in the
    annotations of arcs adjacent to $t$ and data values. The usage of
    pids is restricted to comparisons of the elements from $\Pi_t \cup
    \{ \seq{p}{1}{k} \}$ using the operators from $\{ =, \parent_1,
    \parent, \sibling_1, \sibling \}$
\end{enumerate}

For each reachable marking $M$ we define the corresponding
\emph{state} $q_M \defeq (\sigma_M, \eta_M)$ given by $\sigma_M \defeq
\{ (s, M(s)) \st s \in S \setminus\!\{ \generatorplace \} \}$ and
$\eta_M \defeq \{ \pi \mapsto k \st \tuple{\pi, k} \in
M(\generatorplace) \}$.  Given a state $q_M$ we define the following
notions:
\begin{itemize}
\item for each generative pid $\pi$ in $q_M$, \ie, $\pi \in \dom(\eta_M)$,
  the next pid to be created is given by $\next_{q_M}(\pi) \defeq
  \pi.(\eta_M(\pi)+1)$;
\item the next-pids of $q_M$ are $\nextpid_{q_M} \defeq \{
  \next_{q_M}(\pi) \st \pi \in \dom(\eta_M) \}$;
\item $\pid_{q_M}$ is the set of all pids involved in $\sigma_M$, \ie,
  pids from $\generatorplace$ and all data places.
\end{itemize}

\begin{definition}[state equivalence] Two states
  $q_M$ and $q_{M'}$ are equivalent if there is a bijection $h :
  (\pid_{q_M} \cup \nextpid_{q_M}) \rightarrow (\pid_{q_{M'}} \cup
  \nextpid_{q_{M'}})$ such that for all relations $\prec\;\in \{
  \parent_1, \parent\}$ and $\curlywedge \in \{ \sibling_1, \sibling
  \}$:
    \begin{enumerate}
    \item $h(\dom(\eta_{q_M})) = \dom(\eta_{q_{M'}})$;
    \item $\forall \pi \in \dom(\eta_{q_M}), h(\next_{q_M}(\pi)) =
      \next_{q_{M'}}(h(\pi))$;
    \item $\forall \pi, \pi' \in \pid_{q_M}: \pi \prec \pi'$ iff
      $h(\pi) \prec h(\pi')$;
    \item $\forall \pi, \pi' \in \pid_{q_M} \cup \nextpid_{q_M}: \pi
      \curlywedge \pi'$ iff $h(\pi) \curlywedge h(\pi')$;
    \item $\sigma_{q_{M'}}$ is $\sigma_{q_M}$ after replacing each pid $\pi$
      by $h(\pi)$.
    \end{enumerate}
\noindent We denote this by $q_M \sim_h q_{M'}$, or simply $q_M \sim q_{M'}$.
\end{definition}

Two reachable markings $M$ and $M'$ of a Petri net are equivalent if and
only if their respective states $q_M$ and $q_{M'}$ are
equivalent. This is denoted by $M \sim_h M'$ or simply $M \sim M'$.

This equivalence relation ensures that markings contain the same data
tokens and pids are related through $h$. State (or context)
equivalence guarantees that $h$ preserves the relations between pids for
$\parent_1$, $\parent$, $\sibling_1$ and $\sibling$. So if two
markings are equivalent then they differ in pids but these pids have
the same relations among them and thus differ only in names; names being
not relevant the states can be assimilated.

\def\overtilde#1{ \widetilde{#1} }
\def\compose{\circ}
\begin{theorem} \label{thm::markingequiv}
  Let $M$ and $M'$ be h-equivalent reachable markings of a t-net $N$,
  and $t$ be a transition such that $M\fire{t,
    \beta}\overtilde{M}$. Then $M'\fire{t, h \compose
    \beta}\overtilde{M'}$, where $\overtilde{M'}$ is a marking such
  that $\overtilde{M} \sim_{\overtilde{h}} \overtilde{M'}$ for a
  bijection $\overtilde{h}$ coinciding with $h$ on the intersection of
  their domains.
\end{theorem}

As stated in \cite{KKPP08,KKPP09,KKPP10} the above result captures a
truly strong notion of marking similarity. So, if two markings are
equivalent their futures are the same modulo pid renaming. To compute
state equivalence, the approach proposed in
\cite{KKPP08,KKPP09,KKPP10} was first to build a graph then compute
graph isomorphisms. Computing these isomorphisms is an expensive step
and was hardly conceivable in a tool since we have to compute it for
all visited states each time a new state is discovered. The
contribution of this paper is to provide a representation for markings
to detect this equivalence. This is presented in following sections
but before moving on we add some notations for tokens.

\begin{definition}[owned, shared, active, and referenced pids]
  Let $v = \tuple{x_1, \dots, x_n}$ be a token. If $x_1 \in \set{P}$
  then $x_1$ is the \emph{owner} of $v$, otherwise if $x_1 \in
  \set{D}$ we say that $v$ is \emph{shared}. We denote by $\pid(v)$
  the owner of $v$ if defined. If a pid $\pi$ appears in a token $v$
  at a marking $M$ then we say that $\pi$ is \emph{active} at
  $M$. Finally we define the set of \emph{referenced} pids of $v$ as
  the set of all pids appearing in $v$ except $\pid(v)$.
\end{definition}

\section{Pid-trees}

A pid-tree is a representation of a marking where tokens are
associated to process identifiers. The intention is to classify them
by ownership. The root of the tree will contain data tokens that are
shared between different processes and each of the remaining nodes
will contain tokens that belong to a particular process. To identify
these processes each child node is prefixed with a fragment of
pid. Each path in the tree leads to a node that contains a marking
where all tokens belong to the process which pid is the concatenation
of all the fragment pids along the path; we say that this pid labels
the path. A path can be seen as a concatenation of pids that lead to
a, possibly empty, marking. A formal definition of paths is given
later in this section and ownership constraints will be added when
exposing the construction of the tree. Pid-trees are formally defined
by the following definition.

\begin{definition}[pid-tree] \label{def::pidtree}
  The set $\pidtreeset$ of \emph{pid-trees} is defined by $\tuple{M,
    C} \in \pidtreeset$ if $M \in \markings$ and
   $\exists n \in \set{N}$ such that $C = \tuple{ \tuple{a_1,
       t_1}, \dots, \tuple{a_n, t_n}} \in (\set{P} \times
     \pidtreeset)^{n}$ and $C$ satisfies:
     \begin{itemize}
     \item $\forall i \in \{ 1, \dots, n \}$, $a_i \neq \tuple{}$;
     \item $\forall i, j \in \{ 1, \dots, n \}$ if $i \neq j$ then
       $a_i \neq a_j$, $a_i \nsubseteq \subpid(a_j)$ and $a_j
       \nsubseteq \subpid(a_i)$.
     \end{itemize}
   We denote $M \childarrow{a_1, \dots, a_n} \tuple{t_1, \dots, t_n}$
   the pid-tree $\tuple{ M, \tuple{ \tuple{a_1, t_1}, \dots,
       \tuple{a_n, t_n} } }$.  We will also use the notation $M_t$ to
   denote the marking $M$ in $t = \tuple{ M, C }$.
 \end{definition}

Having tuples in the definition of children nodes allows to perform a
syntactical ordering of them.
The conditions in the second point of definition \ref{def::pidtree}
ensure that each pid associated with a node is unique, but more than
that, it ensures that we cannot have two children prefixed by the same
pid fragment. For example $\emptyset \childarrow{1,1} \tuple{t_1,
  t_2}$ is an illegal pid-tree as well as $\emptyset
\childarrow{1,1.2} \tuple{t_1, t_2}$ since $1 \in \subpid(1.2)$. A
pid-tree can be represented graphically, see Figure
\ref{fig::pidtree}, each node is labelled with a marking and each arc
with a pid fragment. In Figure \ref{fig::pidtree::a}, the process of
pid $1$ does not own any tokens so the corresponding node is
associated to an empty marking, but the process of pid $1.2.1.3$ owns
some tokens so the corresponding node is associated with marking
$M_3$.

\tikzstyle{treenode}=[circle, draw, inner sep=2pt]
\tikzstyle{arc}=[->, draw]

\begin{figure}[tbh]
\centering
\subfigure[]{\label{fig::pidtree::a}
  \begin{tikzpicture}[yscale=-0.9,xscale=1.3]

    \node[treenode,label=right:{$M_1$}]         (root)     at (1, 0) {};
    \node[treenode,label=right:{$\emptyset$}] (node1)    at (1, 1) {};
    \node[treenode,label=right:{$\emptyset$}] (node121)  at (0.5, 2) {};
    \node[treenode,label=right:{$M_2$}] (node11)   at (1.5, 2) {};
    \node[treenode,label=right:{$M_3$}] (node1211) at (0, 3) {};
    \node[treenode,label=right:{$M_4$}] (node111)  at (1, 3) {};
    \node[treenode,label=right:{$\emptyset$}] (node112)  at (2, 3) {};

    \draw [arc] (root)    -- node[right] {$1$} (node1) ;
    \draw [arc] (node1)   -- node[left] {$2.1$} (node121);
    \draw [arc] (node1)   -- node[right] {$1$} (node11);
    \draw [arc] (node121) -- node[left] {$3$} (node1211);
    \draw [arc] (node11)  -- node[left] {$1$} (node111);
    \draw [arc] (node11)  -- node[right] {$2$} (node112);

  \end{tikzpicture}
}\qquad
\subfigure[]{\label{fig::pidtree::b}
  \begin{tikzpicture}[yscale=-0.9,xscale=1.3]
    \node[treenode,label=right:{$M'_1$}]         (root)     at (1.5, 0) {};
    \node[treenode,label=right:{$\emptyset$}]   (node1)    at (1.5, 1) {};
    \node[treenode,label=right:{$\emptyset$}] (node11)   at (1.5, 2) {};
    \node[treenode,label=right:{$M'_4$}] (node111)  at (1, 3) {};
    \node[treenode,label=right:{$\emptyset$}] (node112)  at (2, 3) {};

    \draw [arc] (root)    -- node[right] {$1$} (node1) ;
    \draw [arc] (node1)   -- node[right] {$1$} (node11);
    \draw [arc] (node11)  -- node[left] {$1$} (node111);
    \draw [arc] (node11)  -- node[right] {$2$} (node112);
  \end{tikzpicture}
} \qquad
\subfigure[]{\label{fig::pidtree::c}
  \begin{tikzpicture}[yscale=-0.9,xscale=1.3]

    \node[treenode,label=right:{$\emptyset$}]   (root)     at (1.5, 0) {};
    \node[treenode,label=right:{$\emptyset$}]   (node1)    at (1.5, 1) {};
    \node[treenode,label=right:{$\emptyset$}]   (node11)   at (1.5, 2) {};
    \node[treenode,label=right:{$M'_4$}]        (node111)  at (1.5, 3) {};

    \draw [arc] (root)    -- node[right] {$1$} (node1);
    \draw [arc] (node1)   -- node[right] {$1$} (node11);
    \draw [arc] (node11)  -- node[right] {$1$} (node111);
  \end{tikzpicture}
} \qquad
\subfigure[]{\label{fig::pidtree::d}
  \begin{tikzpicture}[yscale=-0.9,xscale=1.3]

    \node[treenode,label=right:{$\emptyset$}]   (stripped root)     at (2.5, 0) {};
    \node[treenode,label=right:{$M'_4$}]        (stripped node111)  at (2.5, 1) {};

    \node[treenode,label=right:{\color{white}$M'_4$}, draw=white] (phantom)  at (2.5, 3) {};

    \draw [arc] (stripped root)  -- node[right] {$1.1.1$} (stripped node111);
  \end{tikzpicture}
}

\caption{The pid-tree $M_1 \childarrow{1} \tuple{ \emptyset
    \childarrow{2.1, 1} \tuple{ \emptyset \childarrow{3}
      M_3, M_2 \childarrow{1,2} \tuple{M_4, \emptyset} }
  }$ is shown in Figure \ref{fig::pidtree::a}; the pid-tree $M_1
  \childarrow{1} \emptyset \childarrow{1} \emptyset
  \childarrow{1,2} \tuple{M'_4, \emptyset}$ is shown in
  Figure \ref{fig::pidtree::b}; Figures \ref{fig::pidtree::c} and
  \ref{fig::pidtree::d} show two pid-trees that represents
  $\path(1.1.1, M'_4)$ and are respectively $\emptyset \childarrow{1}
  \emptyset \childarrow{1} \emptyset \childarrow{1} M'_4$ and $\emptyset
  \childarrow{1.1.1} M'_4$. We assume that markings $M_i$ and $M'_i$
  are not empty.}
\label{fig::pidtree}
\end{figure}

To check the inclusion of pid-trees we have to start from the root,
first we check the inclusion of markings and then we check the
inclusion of children that must be labelled with the same pid
fragments. It is important to notice that the inclusion operation on
pid-trees preserves the root, respects marking inclusion and does not
denote a subtree relation. The formal definition follows.

\begin{definition}[inclusion] Let $t = \tuple{M, \tuple{ \tuple{a_1, t_1}, \dots,
  \tuple{a_n, t_n}}} \in \pidtreeset$ be a pid-tree. A pid-tree $t' =
  \tuple{M', \tuple{\tuple{a'_1, t'_1}, \dots, \tuple{a'_m, t'_m}}}
  \in \pidtreeset$, with $m \leq n$, is included in $t$, noted $t'
  \subseteq t$, if $M'\le M$ and for all $i \in \{1, \dots, m\}$,
  there is a $j \in \{1, \dots, n\}$ such that $a'_i = a_j$ and $t'_i
  \subseteq t_j$.
\end{definition}
For example, the pid-trees in Figure \ref{fig::pidtree::b} and
\ref{fig::pidtree::c} are included in the pid-tree in Figure
\ref{fig::pidtree::a} if $M'_1 \le M_1$ and $M'_4 \le M_4$. The
pid-tree in Figure \ref{fig::pidtree::d} is not included is any other
pid-trees in Figure \ref{fig::pidtree}. The pid-tree $\emptyset
\childarrow{1} \emptyset$ is included in pid-trees in Figures
\ref{fig::pidtree::a}, \ref{fig::pidtree::b} and \ref{fig::pidtree::c}
whereas $\emptyset \childarrow{2} \emptyset$ is not included in any of
pid-trees in Figure \ref{fig::pidtree}, which is straightforward to
understand if we remember that we have to preserve the root.

The definition of subtrees is similar to the usual definition on
trees, however we add a localisation hint for the subtree in order to
differentiate two subtrees that are structurally equal but at
different positions in the tree. Thus a subtree will be a pair formed
of a pid and a pid-tree.

\begin{definition}[subtrees]
  Let $t_0 = M \childarrow{a_1, \dots, a_n} \tuple{ t_1, \dots, t_n }
  \in \pidtreeset$ be a pid-tree. A pid-tree $t$ is a subtree of $t_0$
  at $\pi$, noted $\tuple{\pi, t} \in \trees(t_0)$, if:
  \begin{itemize}
    \item $\pi = \tuple{}$ and $t = t_0$; or
    \item $\exists a_i \in \{ a_1, \dots, a_n \}$ such that $\pi = a_i.\pi'$ and $\tuple{\pi', t} \in \trees(t_i)$.
  \end{itemize}
  We denote $\trees(t\!)$ the set of all subtrees of $t$.
\end{definition}

Using the definition of subtrees, we can retrieve the set of all pids
in a pid-tree which corresponds to all the localisations in the
pid-tree.

\begin{definition} Let $t \in \pidtreeset$ be a pid-tree. We define
  $\pid(t) = \{~ \pi ~|~ \tuple{\pi, t'} \in \trees(t) ~\}$.
\end{definition}

Now that we have set up some basic blocks to handle pid-trees, we give
a formal definition of paths. A path is a \emph{linear} pid-tree with
all markings empty except for the leaf one.

\begin{definition}[paths] \label{def::decoratedpath}
  Let $\pi \in \set{P}$ be a pid, and $M \in \markings$ a marking. A
  \emph{path} labeled by $\pi$ and decorated by $M$ is a pid-tree $t$
  such that $\{ \pi \} \subseteq \pid(t) \subseteq \subpid(\pi)$,
  $\tuple{\pi, \tuple{M, \tuple{}}} \in \trees(t)$ and all other
  markings in $t$ are empty. We denote $\path(\pi)$ the path
  $\path(\pi, \emptyset)$ and $\paths(T)$ the set of all paths in a
  pid-tree $T$(with respect to $\subseteq$).
\end{definition}

As the definition suggests, there may be different pid-tree
representations of a path, indeed these representations will differ in
the length of the path and the pid fragments on arcs. For example the
pid-trees on Figures \ref{fig::pidtree::c} and \ref{fig::pidtree::d}
represents the same path $\path(1.1.1, M'_4)$.

A path denotes an ownership relation between a process and some
tokens. Intuitively $\path(\pi, M)$ represents the fact that all
tokens that appear in $M$ belong to the process of pid $\pi$. We say
that $\pi$ labels this path. Moreover saying that a subtree $t$ is at
$\pi$ means that the path $path(\pi)$ leads to $t$.

When addressing process identifiers, their names are irrelevant, so we
introduce an opaque description for localisations in pid-trees. We
could use pids to locate nodes but we need to abstract them so we use
relative paths which describe the localisation of a subtree
independently of the labels appearing on the edges. A relative path
can be seen as a normalized form of a pid based on its position inside
the tree.

\begin{definition}[relative paths] Let $t \in \pidtreeset$ be a
  pid-tree. A relative \emph{path} $\relpath{\pi, t}$ of $\pi \in
  \pid(t)$ is the tuple defined as:
  \begin{itemize}
  \item $\tuple{i}$ if $\pi = a_i$ and $t = M \childarrow{a_1, \dots, a_n}
    \tuple{t_1, \dots, t_n}$ where $1 \le i \le n$;
  \item $\tuple{i, i_1, \dots, i_m}$ if $\pi = a_i . \pi'$,
    $t = M \childarrow{a_1, \dots, a_n} \tuple{t_1, \dots, t_n}$ and
    $\tuple{i_1, \dots, i_m} = \relpath{\pi', t_i}$ where $1
    \le i \le n$.
  \end{itemize}
\end{definition}

Finally, since we have the ability to consider the ordering of
children of a node, we introduce labelling functions that will be used
to compare and order these children.

\begin{definition} Let $(E, \leq)$ be a partially ordered set.
  A labelling function $\ell$ is a partial function from $\set{P}
  \times \pidtreeset$ to $ E $.
\end{definition}

\begin{definition} Let $t \in \pidtreeset$ be a pid-tree and $\ell$ a
  labelling function. The pid-tree $t$ is ordered with respect to
  $\ell$ if:
  \begin{itemize}
  \item $t = \tuple{M, \tuple{}}$ or
  \item $t = M \childarrow{a_1, \dots, a_n} \tuple{t_1, \dots, t_n}$, 
    $\ell(a_1, t_1) \leq \dots \leq \ell(a_n, t_n)$ and
    $\seq{t}{1}{n}$ are ordered with respect to $\ell$.
  \end{itemize}
\end{definition}

\section{Checking marking equivalence}

To apply theorem \ref{thm::markingequiv} and reduce state spaces we
will need to detect marking equivalence. We check marking equivalence
by mapping markings to pid-trees and comparing these trees. If the
trees are equivalent, markings will be isomorphic. However the
representation we use is tightly linked to the relations used to
compare pids in the models. More precisely the representation we use
embeds the relations $\parent_1$, $\parent$, $\sibling_1$ and
$\sibling$.

Intuitively a pid-tree embeds the relation $\parent$ by construction
because it is a hierarchy of pids. In this section we need to include the
relation $\sibling$ into pid-trees as well as $\parent_1$ and
$\sibling_1$. To do so, we will use a specific labelling and ordering
on child nodes.

\begin{definition}[sibling ordering] Let $t \in
  \pidtreeset$ be a pid-tree. The pid-tree $t$ is \emph{sibling
    ordered} if it is ordered with the labelling function $\ell :
  \set{P} \times \pidtreeset \rightarrow \set{P}$ defined as:
  $$
  \ell(a_i,\;t') = a_i
  $$
  Where $\set{P}$ is equipped with the hierarchical order on pids.
\end{definition}

Using the sibling ordering we can define the representation of
markings that will be used to detect symmetries. The representation is
described in terms of three rules that have to be obeyed when building
the pid-tree. The resulting pid-tree is not unique, as shown in Figure
\ref{fig::reprM}, but the definition gives all the theoretical
requirements to guarantee the marking equivalence.

\def\repr{\mathit{repr}}
\begin{definition}[pid-tree representation of markings]
  \label{def::pidtreerepresentation} Let $M$ be a
  reachable marking of a t-net $N$. We build a representation of $M$
  as a sibling ordered pid-tree and $\repr(M)$ denotes the set of all
  such representations. $R(M) \in \repr(M)$ if for each place $s \in
  N$ which type is $X_1 \times \dots \times X_n$, and for each token
  $v = \tuple{x_1, \dots, x_n} \in M(s)$, for $X_i \in \{ \set{D},
  \set{P} \}$, we apply exactly one of the following rules:
  \begin{itemize}
  \item \emph{\textbf{generator token rule:}} if $s = \generatorplace$
    and $v = \tuple{\pi, i}$ then we have $\path(\pi) \subseteq R(M)$
    and $\path(\pi.(i+1)) \subseteq R(M)$, \ie,
    $\path(\next_{q_M}(\pi)) \subseteq R(M)$.
  \item \emph{\textbf{shared token rule:}} if $X_1 = \set{D}$ then we
    have $\path(\tuple{}, \{ (s, v) \}) \subseteq R(M)$, and for each
    $x_i$ such that $X_i = \set{P}$ we have $\path(x_i) \subseteq R(M)$.
  \item \emph{\textbf{owned token rule:}} if $X_1 = \set{P}$ then we
    have $\path(x_1, \{ (s, v) \}) \subseteq R(M)$, and for each
    $x_i$ such that $X_i = \set{P}$ we have $\path(x_i) \subseteq R(M)$.
  \end{itemize}
\end{definition}

\begin{figure}[htb]
\centering
\subfigure[]{
  \begin{tikzpicture}[yscale=-1,xscale=1.3]

    \node[treenode,label=right:{$\{ s_1 \mapsto \{ 12 \} \}$}]         (root)     at (1, 0) {};
    \node[treenode,label=right:{$\{ s_2 \mapsto \{ \tuple{1} \} \}$}] (node1)    at (1, 1) {};
    \node[treenode,label=right:{$\emptyset$}] (node121)  at (0.5, 2) {};
    \node[treenode,label=right:{$\emptyset$}] (node11)   at (1.5, 2) {};
    \node[treenode,label=right:{$\{ s_2 \mapsto \{ \tuple{1,1,2} \} \}$}] (node1211) at (0.5, 3) {};
    \node[treenode,label=right:{$\emptyset$}] (node12111) at (0.5, 4) {};

    \draw [arc] (root)    -- node[right] {$1$} (node1) ;
    \draw [arc] (node1)   -- node[left] {$1$} (node121);
    \draw [arc] (node1)   -- node[right] {$2$} (node11);
    \draw [arc] (node121) -- node[left] {$2$} (node1211);
    \draw [arc] (node1211) -- node[left] {$1$} (node12111);

  \end{tikzpicture}
}\qquad
\subfigure[]{
  \begin{tikzpicture}[yscale=-1,xscale=1.3]

    \node[treenode,label=right:{$\{ s_1 \mapsto \{ 12 \} \}$}]         (root)     at (1, 0) {};
    \node[treenode,label=right:{$\{ s_2 \mapsto \{ \tuple{1} \} \}$}] (node1)    at (1, 1) {};
    \node[treenode,label=right:{$\{ s_2 \mapsto \{ \tuple{1,1,2} \}$}] (node121)  at (1.5, 2) {};
    \node[treenode,label=right:{$\emptyset$}] (node11)   at (0.5, 2) {};
    \node[treenode,label=right:{$\{ \emptyset \}$}] (node1211) at (1.5, 3) {};
    \node[treenode,white,label=right:{\color{white}$\{ \emptyset \}$}] at (1.5, 4) {};

    \draw [arc] (root)    -- node[right] {$1$} (node1) ;
    \draw [arc] (node1)   -- node[right] {$1.2$} (node121);
    \draw [arc] (node1)   -- node[left] {$2$} (node11);
    \draw [arc] (node121) -- node[right] {$1$} (node1211);

  \end{tikzpicture}
}
\vspace{-6pt}

\caption{Two sibling ordered pid-trees representing the marking $M = \{
  s_1 \mapsto \{12\}, s_2 \mapsto \{ \tuple{1}, \tuple{1,1,2} \},
  \generatorplace \mapsto \{ \tuple{1, 1}, \tuple{ \tuple{1,1,2}, 0 } \}
  \}$ with $\ell(s_1) = \set{N}$, $\ell(s_2) = \set{P}$ and
  $\ell(\generatorplace) = \set{P} \times \set{N}$.}
\label{fig::reprM}

\end{figure}

Checking the equality on sibling ordered pid-trees will not offer any
advantages over comparing markings for symmetry detection. To capture
symmetries we need to abstract pids because their values are irrelevant
and only the relations between pids matter. This leads to the
definition of an equivalence relation of sibling ordered pid-trees.

\begin{definition}[sibling ordered pid-tree
    equivalence] \label{def::siblingequiv}
  Let $T\!$ and $T'\!$ be two sibling ordered pid-trees, $T\!$ and $T'\!$ are
  equivalent if:
  \begin{enumerate}
  \item the function $h \,:\, \pid(T) \rightarrow \pid(T')$ such that $ h
    = \{ ~ (\pi, \pi') ~~|~~ \pi \in \pid(T),~~ \pi' \in \pid(T')
    ~\mbox{ and } \newline ~ \relpath{\pi, T} \,=\, \relpath{\pi', T'} ~ \} $ is a
    bijection
    \item for each pair of subtrees $\tuple{\pi, t } \in \trees(T)$
      and $\tuple{\pi', t' } \in \trees(T')$ such that $\pi' = h(\pi)$
      and $t = M_t \childarrow{a_1, \dots, a_n} \tuple{t_1, \dots,
        t_n}$, $t' = M_{t'} \childarrow{a'_1, \dots, a'_{n'}}
      \tuple{t'_1, \dots, t'_{n'}}$ we have:
      \begin{enumerate}
      \item trees have the same structure, \ie, $n = n'$, which is implied by the definition of $h$.
      \item for each place $s$ different from $\generatorplace$, $M_{t'}(s)$ can be
        obtained from $M_t(s)$ by replacing each pid $\pi$ occurring
        in the tuples of $M_t(s)$ by $h(\pi)$;
      \item for each $a_i$ we have
        \begin{itemize}
          \item $\length(a_i) = \length(a'_i) = 1$; or
          \item $\length(a_i) > 1$ and $\length(a'_i) > 1$.
        \end{itemize}
      \item for each $a_i$ and $a_{i+1}$ such that
        $\prefix(a_i) = \prefix(a_{i+1})$ and $1 \le i \le n-1$ we
        have $\prefix(a'_i) = \prefix(a'_{i+1})$ and
        \begin{itemize}
        \item $a_{i+1} - a_i = a'_{i+1} - a'_i = 1$; or
        \item $a_{i+1} - a_i > 1$ and $a'_{i+1} - a'_i > 1$.
        \end{itemize}
      \end{enumerate}
  \end{enumerate}
  We will denote this by $T \siblingequivh{h} T'$ or simply by $T
  \siblingequiv T'$.
\end{definition}

The function $h$ in the definition above is a bijection if and only if
two pid-trees have the same structure because pids are related through
$h$ if and only if their relative paths are equal. The second point
states first that trees have the same structure, the markings only
differ in pids and these pids are related through $h$, then checks
condition on pid fragments. Length conditions ensure that $\parent_1$
and $\parent$ are preserved, indeed if the length of a fragment is
equal to one then the parent and the child are in $\parent_1$
otherwise the parent and the child are in $\parent$. Offset conditions
for fragments with same prefix ensure that $\sibling_1$ and $\sibling$
are preserved. If prefixes are equal then the children are
siblings. If offsets are equal to one then the corresponding pids are
in $\sibling_1$ and if offsets are greater than one the pids are in
$\sibling$.

It has to be stressed that we cannot use $\length(a_i) =
\length(a'_i)$ instead of $\length(a_i) > 1$ and $\length(a'_i) > 1$
because the the number of inactive pids between them and their parents
does not matter as long as $\parent$ is preserved. We qualify a pid as
\emph{inactive} if it does not appear in the marking (including the
generator place) used to build the tree. The same remark applies for
offsets and $\sibling$.

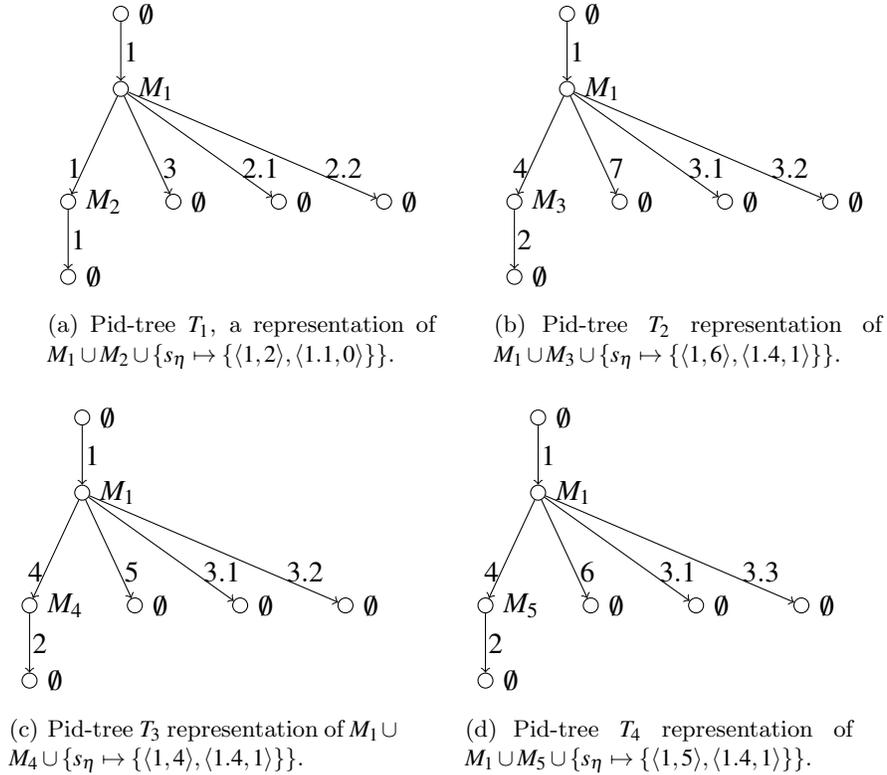
\begin{figure}[bth]
  \centering
\subfigure[Pid-tree $T_1$, a representation of $M_1 \cup M_2 \cup \{ \generatorplace
  \mapsto \{ \tuple{1, 2}, \tuple{1.1, 0} \} \}$.]
{\label{fig::EquivM::a}
  \begin{tikzpicture}[yscale=-1,xscale=1.4]

    \node[treenode,label=right:{$\emptyset$}] (root)      at (0.5, 0) {};
    \node[treenode,label=right:{$M_1$}] (node1)     at (0.5, 1) {};
    \node[treenode,label=right:{$M_2$}] (node11)    at (0, 2.5) {};
    \node[treenode,label=right:{$\emptyset$}] (node13)    at (1, 2.5) {};
    \node[treenode,label=right:{$\emptyset$}] (node121)   at (2, 2.5) {};
    \node[treenode,label=right:{$\emptyset$}] (node122)   at (3, 2.5) {};
    \node[treenode,label=right:{$\emptyset$}] (node111)   at (0, 3.5) {};

    \draw [arc] (root)    -- node[right=-0.25em] {$1$} (node1) ;
    \draw [arc] (node1)   -- node[left=0.7em,below=0.1em]    {$1$}   (node11);
    \draw [arc] (node1)   -- node[right=0.8em, below=0.1em] {$3$}   (node13);
    \draw [arc] (node1)   -- node[right=2.1em,below=0.1em]  {$2.1$} (node121);
    \draw [arc] (node1)   -- node[right=3.15em,below=0.1em] {$2.2$} (node122);
    \draw [arc] (node11)  -- node[right=-0.25em] {$1$} (node111);

  \end{tikzpicture}
}\qquad
\subfigure[Pid-tree $T_2$ representation of
  $M_1 \cup M_3 \cup \{ \generatorplace
  \mapsto \{ \tuple{1, 6}, \tuple{1.4, 1} \} \}$.
]{\label{fig::EquivM::b}
  \begin{tikzpicture}[yscale=-1,xscale=1.4]

    \node[treenode,label=right:{$\emptyset$}] (root)      at (0.5, 0) {};
    \node[treenode,label=right:{$M_1$}] (node1)     at (0.5, 1) {};
    \node[treenode,label=right:{$M_3$}] (node11)    at (0, 2.5) {};
    \node[treenode,label=right:{$\emptyset$}] (node13)    at (1, 2.5) {};
    \node[treenode,label=right:{$\emptyset$}] (node121)   at (2, 2.5) {};
    \node[treenode,label=right:{$\emptyset$}] (node122)   at (3, 2.5) {};
    \node[treenode,label=right:{$\emptyset$}] (node111)   at (0, 3.5) {};

    \draw [arc] (root)    -- node[right=-0.25em] {$1$} (node1) ;
    \draw [arc] (node1)   -- node[left=0.7em,below=0.1em]    {$4$}   (node11);
    \draw [arc] (node1)   -- node[right=0.8em, below=0.1em] {$7$}   (node13);
    \draw [arc] (node1)   -- node[right=2.1em,below=0.1em]  {$3.1$} (node121);
    \draw [arc] (node1)   -- node[right=3.15em,below=0.1em] {$3.2$} (node122);
    \draw [arc] (node11)  -- node[right=-0.25em] {$2$} (node111);

  \end{tikzpicture}
}
\subfigure[Pid-tree $T_3$ representation of
  $M_1 \cup M_4 \cup \{ \generatorplace
  \mapsto \{ \tuple{1, 4}, \tuple{1.4, 1} \} \}$.
]{\label{fig::EquivM::c}
  \begin{tikzpicture}[yscale=-1,xscale=1.4]

    \node[treenode,label=right:{$\emptyset$}] (root)      at (0.5, 0) {};
    \node[treenode,label=right:{$M_1$}] (node1)     at (0.5, 1) {};
    \node[treenode,label=right:{$M_4$}] (node11)    at (0, 2.5) {};
    \node[treenode,label=right:{$\emptyset$}] (node13)    at (1, 2.5) {};
    \node[treenode,label=right:{$\emptyset$}] (node121)   at (2, 2.5) {};
    \node[treenode,label=right:{$\emptyset$}] (node122)   at (3, 2.5) {};
    \node[treenode,label=right:{$\emptyset$}] (node111)   at (0, 3.5) {};

    \draw [arc] (root)    -- node[right=-0.25em]            {$1$}   (node1) ;
    \draw [arc] (node1)   -- node[left=0.7em,below=0.1em]   {$4$}   (node11);
    \draw [arc] (node1)   -- node[right=0.8em, below=0.1em] {$5$}   (node13);
    \draw [arc] (node1)   -- node[right=2.1em,below=0.1em]  {$3.1$} (node121);
    \draw [arc] (node1)   -- node[right=3.15em,below=0.1em] {$3.2$} (node122);
    \draw [arc] (node11)  -- node[right=-0.25em]            {$2$}   (node111);

  \end{tikzpicture}
} \qquad
\subfigure[Pid-tree $T_4$ representation of
  $M_1 \cup M_5 \cup \{ \generatorplace
  \mapsto \{ \tuple{1, 5}, \tuple{1.4, 1} \} \}$.]{\label{fig::EquivM::d}
  \begin{tikzpicture}[yscale=-1,xscale=1.4]

    \node[treenode,label=right:{$\emptyset$}] (root)      at (0.5, 0) {};
    \node[treenode,label=right:{$M_1$}] (node1)     at (0.5, 1) {};
    \node[treenode,label=right:{$M_5$}] (node11)    at (0, 2.5) {};
    \node[treenode,label=right:{$\emptyset$}] (node13)    at (1, 2.5) {};
    \node[treenode,label=right:{$\emptyset$}] (node121)   at (2, 2.5) {};
    \node[treenode,label=right:{$\emptyset$}] (node122)   at (3, 2.5) {};
    \node[treenode,label=right:{$\emptyset$}] (node111)   at (0, 3.5) {};

    \draw [arc] (root)    -- node[right=-0.25em] {$1$} (node1) ;
    \draw [arc] (node1)   -- node[left=0.7em,below=0.1em]    {$4$}   (node11);
    \draw [arc] (node1)   -- node[right=0.8em, below=0.1em] {$6$}   (node13);
    \draw [arc] (node1)   -- node[right=2.1em,below=0.1em]  {$3.1$} (node121);
    \draw [arc] (node1)   -- node[right=3.15em,below=0.1em] {$3.3$} (node122);
    \draw [arc] (node11)  -- node[right=-0.25em] {$2$} (node111);

  \end{tikzpicture}
} \qquad
\caption{Four sibling ordered pid-trees where
  $M_1 = \{ s_1 \mapsto \{ \tuple{1} \} \}$,
  $M_2 = \{ s_2 \mapsto \{ \tuple{1.1, 2.1, 2.2} \} \}$,
  $M_3 = \{ s_2 \mapsto \{ \tuple{1.4, 3.1, 3.2} \} \}$,
  $M_4 = \{ s_2 \mapsto \{ \tuple{1.4, 3.1, 3.2} \} \}$ and
  $M_5 = \{ s_2 \mapsto \{ \tuple{1.4, 3.1, 3.3} \} \}$.
}
\label{fig::EquivM}
\end{figure}

The equivalence is illustrated in Figure \ref{fig::EquivM}. Pid-trees
in this figure are valid representations of markings, however we do
not explain here how we choose the pid-tree we want among valid
representations, this is discussed in section
\ref{sec::discussion}. Pid-trees $T_1$ (Figure \ref{fig::EquivM::a})
and $T_2$ (Figure \ref{fig::EquivM::b}) are equivalent if $h = \{ 1
\mapsto 1,~ 1.1 \mapsto 1.4,~ 1.3 \mapsto 1.7,~ 1.2.1 \mapsto 1.3.1,~
1.2.2 \mapsto 1.3.2,~ 1.1.1 \mapsto 1.4.2 \}$ satisfies marking,
length and prefix conditions. Marking conditions are satisfied:
$h(\tuple{1}) = \tuple{1}$ and $h(\tuple{1.1,~ 1.2.1,~ 1.2.2}) =
\tuple{1.4,~ 1.3.1,~ 1.3.2}$. Prefix and length conditions are
satisfied as well:
\begin{itemize}
\item we have $\prefix(1) = \prefix(3) = \tuple{}$, $\prefix(4) = \prefix(7) =
  \tuple{}$, we check offsets: $3-1 > 1$ and $7-4 > 1$.
\item we have $\prefix(2.1) = \prefix(2.2) = \tuple{2}$, $\prefix(3.1) = \prefix(3.2) =
  \tuple{3}$, we check offsets: $2-1 = 2-1 = 1$.
\end{itemize}
Pid-trees $T_3$ (Figure \ref{fig::EquivM::c}) and $T_4$ (Figure
\ref{fig::EquivM::d}) are not equivalent with others in Figure
\ref{fig::EquivM} despite they have the same structure, prefix
conditions can be checked to convince ourselves.

Intuitively the length conditions guaranties that $\parent_1$ is
preserved, and the prefix conditions guaranties that the distances
between pids with respect to relation $\sibling_1$ are correct and
thus replaced pids have same relations with their siblings and
ancestors.

Now we will see that we can define a restriction of $h$ to $\pid_{q_M}
\cup \nextpid_{q_M}$ that is also a bijection and that we can extract
the original marking from the pid-tree representation. This is denoted
by the following two propositions. First we bound the pids appearing
in a representation, then we state that different restrictions are
bijections as well.

\def\for{\mbox{ \mathit{for} }}
\begin{proposition} \label{prop::pidbounds}
  Let $M$ be a reachable marking of a t-net, $q_M$ its state and
  $R(M)$ its representation. Then
  $$
  \pid_{q_M} \cup \nextpid_{q_M} \subseteq \pid(R(M)) \subseteq \subpid_{q_M} \cup \nextpid_{q_M}
  $$
  Where $\subpid_{q_M} = \{ \subpid(\pi) \st \pi \in \pid_{q_M} \}$.
\end{proposition}

\begin{proposition} \label{prop::restrictionh}
  Let $M$ and $M'$ be two reachable markings of a t-net, $q_M, q_{M'}$ their
  respective states and $R(M), R(M')$ their representations. If $R(M)
  \siblingequivh{h} R(M')$ then:
  \begin{enumerate}
  \item the restriction $h_{1} : \pid_{q_M} \rightarrow \pid_{q_{M'}}$ of $h$ is a bijection;
  \item the restriction $h_{2} : \nextpid_{q_M} \rightarrow \nextpid_{q_{M'}}$ of $h$ is a bijection.
  \item the restriction $h_{3} : \pid_{q_M} \cup \nextpid_{q_M}
    \rightarrow \pid_{q_{M'}} \cup \nextpid_{q_{M'}}$ of $h$ is a bijection.
  \end{enumerate}
\end{proposition}

The following proposition links all tokens in marking (except tokens
from $\generatorplace$) with tokens in the tree.

\begin{proposition} \label{prop::markingequnionmarkings}
  Let $M$ be a reachable marking of a t-net $N$. Then for each place
  $s$ in $N$ such that $s \neq \generatorplace$ we have:
  $$
  M(s) = \left( \bigcup_{\tuple{\pi, t} \in \trees(R(M))} M_t(s) \right)
  $$
\end{proposition}

The next theorem states that two equivalent representations of
markings imply an equivalence of markings. It leads to potential
reduction in state spaces due to symmetries considering theorem
\ref{thm::markingequiv}.

\begin{theorem} \label{thm::equivimpequiv}
  Let $M_1$ and $M_2$ be two reachable markings of a
  t-net. If $R(M_1)$ and $R(M_2)$ are equivalent
  then $M_1 \sim M_2$.
\end{theorem}

\def\offset{\mathit{offset}}

It is worth noting that at this stage, markings do not have a unique
representation. However in the following section we will introduce a
sufficient condition to detect all symmetries within a system.

\section{Discussing canonisation} \label{sec::discussion}

As shown in previous section, there are different representations of a
marking. These representations differ in the number of paths they
contain. But as shown in proposition \ref{prop::pidbounds} the number
of paths in a pid-tree can be bounded by the following equation:
$$
\pid_{q_M} \cup \nextpid_{q_M}
\subseteq \pid(R(M))
\subseteq \subpid_{q_M} \cup \nextpid_{q_M}
$$

In next two subsections we will explore two canonical representations
of sibling ordered pid-trees. The first corresponding to the upper bound
of the representation and the second one to the lower bound.

\subsection{Expanded pid-tree representation}

At this point we have a representation of markings as sibling ordered
pid-trees and a pid-tree equivalence relation that implies an
equivalence of markings which lead to the detection of
symmetries. However using the rules we provide one can build different
pid-trees that respect the definition
\ref{def::pidtreerepresentation}, indeed the definition of a path
(definition \ref{def::decoratedpath}) allows to strip or expand
branches inside the tree.

\begin{definition}[expanded pid-trees] A pid-tree $T \in
  \pidtreeset$ is expanded if for all $\tuple{\pi, t}$ in $\trees(T)$
  such that $t = M \childarrow{a_1, \dots, a_n} \tuple{t_1, \dots, t_n}$ we have
  $\length(a_i) = 1$ for $1 \le i \le n$.
\end{definition}

In an expanded pid-tree all pids associated to children nodes are of
length $1$. Intuitively we can see an expanded pid-tree as the biggest
representation of a pid-tree, no more nodes can be added to the tree
without breaking the property about bounds. Pid-trees in Figures
\ref{fig::pidtree::b} and \ref{fig::pidtree::c} are expanded pid-trees
but pid-trees in Figures \ref{fig::pidtree::a} and
\ref{fig::pidtree::d} are not.

\begin{proposition} \label{prop::expandunique}
  Let $M$ be a reachable marking of a t-net $N$. There is a unique expanded
  pid-tree in $\repr(M)$.
\end{proposition}

\subsection{Stripped pid-tree representation}

Now we will minimize the tree by removing all inactive pids. When a
path labelled by $\pi$ is added to the tree we potentially add all the
subpids of $\pi$ which may be inactive. Removing all inactive pids
lead to stripped pid-trees.

\begin{definition} A stripped pid-tree is a pid-tree $T$ such that for
  all $\path(\pi) \in \paths(T)$ we have $\pi$ active.
\end{definition}

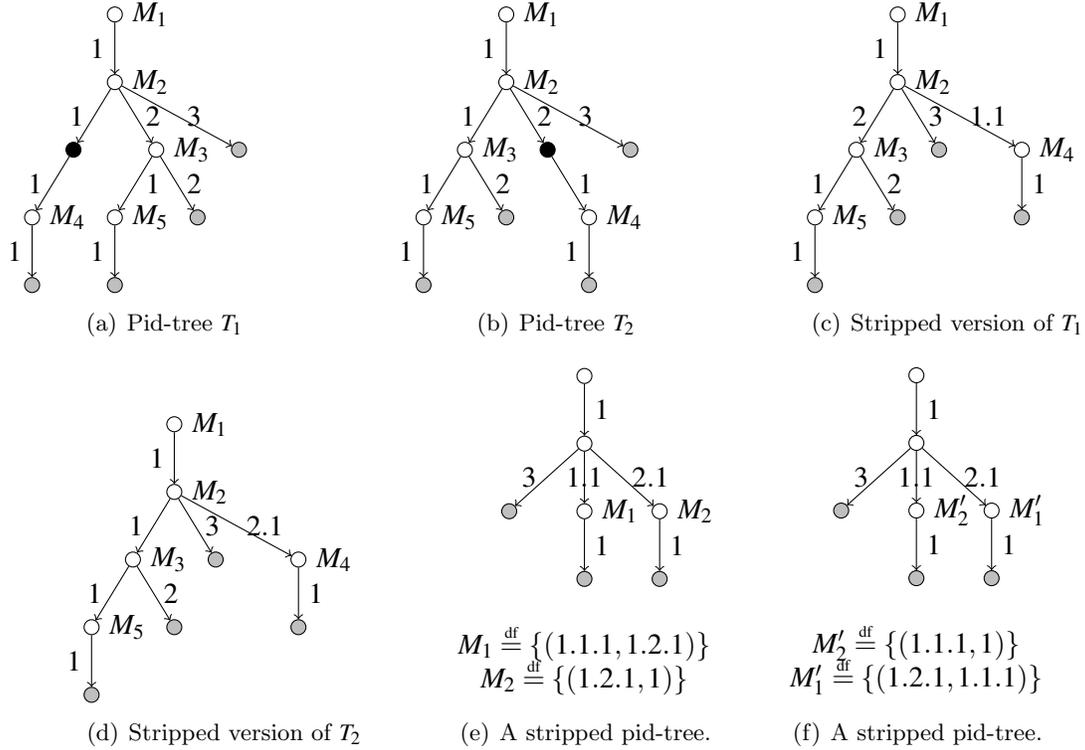
\begin{figure}[bth]
  \centering
\subfigure[Pid-tree $T_1$]
{\label{fig::stripped::a}
  \begin{tikzpicture}[yscale=-0.9,xscale=1.1]

    \node[treenode,label=right:{$M_1$}] (root) at (0.5, 0.5) {};
    \node[treenode,label=right:{$M_2$}] (node1) at (0.5, 1.5) {};
    \node[treenode,label=right:{}, fill=black] (node11) at (0, 2.5) {};
    \node[treenode,label=right:{$M_3$}] (node12) at (1, 2.5) {};
    \node[treenode,label=right:{}, fill=lightgray] (node13) at (2,
    2.5) {};
    \node[treenode,label=right:{$M_4$}] (node111) at (-0.5, 3.5) {};
    \node[treenode,label=right:{$M_5$}] (node121) at (0.5, 3.5) {};
    \node[treenode,label=right:{}, fill=lightgray] (node122) at (1.5,
    3.5) {}; 
    \node[treenode,label=right:{}, fill=lightgray] (node1111) at
    (-0.5, 4.5) {};
    \node[treenode,label=right:{}, fill=lightgray] (node1211) at (0.5, 4.5) {};
    \node at (3, 4.5) {};

    \draw [arc] (root)   -- node[left]  {$1$} (node1);
    \draw [arc] (node1)  -- node[left]  {$1$} (node11);
    \draw [arc] (node1)  -- node[right] {$2$} (node12);
    \draw [arc] (node1)  -- node[right] {$3$} (node13);
    \draw [arc] (node11) -- node[left]  {$1$} (node111);
    \draw [arc] (node12) -- node[right] {$1$} (node121);
    \draw [arc] (node12) -- node[right] {$2$} (node122);
    \draw [arc] (node111) -- node[left] {$1$} (node1111);
    \draw [arc] (node121) -- node[left] {$1$} (node1211);
  \end{tikzpicture}
} \quad
\subfigure[Pid-tree $T_2$]
{\label{fig::stripped::b}
  \begin{tikzpicture}[yscale=-0.9,xscale=1.1]

    \node[treenode,label=right:{$M_1$}] (root)      at (0.5, 0.5) {};
    \node[treenode,label=right:{$M_2$}] (node1)     at (0.5, 1.5) {};
    \node[treenode,label=right:{$M_3$}] (node11)    at (0, 2.5) {};
    \node[treenode,label=right:{}, fill=black] (node12)    at (1, 2.5) {};
    \node[treenode,label=right:{}, fill=lightgray] (node13)   at (2, 2.5) {};
    \node[treenode,label=right:{$M_5$}] (node111)   at (-0.5, 3.5) {};
    \node[treenode,label=right:{}, fill=lightgray] (node112)   at (0.5, 3.5) {};
    \node[treenode,label=right:{$M_4$}] (node121)   at (1.5, 3.5) {};
    \node[treenode,label=right:{}, fill=lightgray] (node1111)  at (-0.5, 4.5) {};
    \node[treenode,label=right:{}, fill=lightgray] (node1211)  at (1.5, 4.5) {};

    \node at (3, 4.5) {};

    \draw [arc] (root)   -- node[left]  {$1$} (node1);
    \draw [arc] (node1)  -- node[left]  {$1$} (node11);
    \draw [arc] (node1)  -- node[right] {$2$} (node12);
    \draw [arc] (node1)  -- node[right] {$3$} (node13);
    \draw [arc] (node11) -- node[left]  {$1$} (node111);
    \draw [arc] (node11) -- node[right] {$2$} (node112);
    \draw [arc] (node12) -- node[right] {$1$} (node121);
    \draw [arc] (node111) -- node[left] {$1$} (node1111);
    \draw [arc] (node121) -- node[left] {$1$} (node1211);
  \end{tikzpicture}
} \quad
\subfigure[Stripped version of $T_1$]
{\label{fig::stripped::c}
  \begin{tikzpicture}[yscale=-0.9,xscale=1.1]

    \node[treenode,label=right:{$M_1$}] (root)      at (0.5, 0.5) {};
    \node[treenode,label=right:{$M_2$}] (node1)     at (0.5, 1.5) {};
    \node[treenode,label=right:{$M_3$}] (node11)    at (0, 2.5) {};
    \node[treenode,label=right:{$M_4$}] (node12)    at (2, 2.5) {};
    \node[treenode,label=right:{}, fill=lightgray] (node13)   at (1, 2.5) {};
    \node[treenode,label=right:{$M_5$}] (node111)   at (-0.5, 3.5) {};
    \node[treenode,label=right:{}, fill=lightgray] (node112)   at (0.5, 3.5) {};
    \node[treenode,label=right:{}, fill=lightgray] (node121)   at (2, 3.5) {};
    \node[treenode,label=right:{}, fill=lightgray] (node1111)  at (-0.5, 4.5) {};

    \node at (3, 4.5) {};

    \draw [arc] (root)   -- node[left]  {$1$} (node1);
    \draw [arc] (node1)  -- node[left]  {$2$} (node11);
    \draw [arc] (node1)  -- node[right] {$1.1$} (node12);
    \draw [arc] (node1)  -- node[right] {$3$} (node13);
    \draw [arc] (node11) -- node[left]  {$1$} (node111);
    \draw [arc] (node11) -- node[right] {$2$} (node112);
    \draw [arc] (node12) -- node[right] {$1$} (node121);
    \draw [arc] (node111) -- node[left] {$1$} (node1111);
  \end{tikzpicture}
} \quad
\subfigure[Stripped version of $T_2$]
{\label{fig::stripped::d}
  \begin{tikzpicture}[yscale=-0.9,xscale=1.1]

    \node[treenode,label=right:{$M_1$}] (root)      at (0.5, 0.5) {};
    \node[treenode,label=right:{$M_2$}] (node1)     at (0.5, 1.5) {};
    \node[treenode,label=right:{$M_3$}] (node11)    at (0, 2.5) {};
    \node[treenode,label=right:{$M_4$}] (node12)    at (2, 2.5) {};
    \node[treenode,label=right:{}, fill=lightgray] (node13)   at (1, 2.5) {};
    \node[treenode,label=right:{$M_5$}] (node111)   at (-0.5, 3.5) {};
    \node[treenode,label=right:{}, fill=lightgray] (node112)   at (0.5, 3.5) {};
    \node[treenode,label=right:{}, fill=lightgray] (node121)   at (2, 3.5) {};
    \node[treenode,label=right:{}, fill=lightgray] (node1111)  at (-0.5, 4.5) {};

    \node at (3, 4.5) {};

    \draw [arc] (root)   -- node[left]  {$1$} (node1);
    \draw [arc] (node1)  -- node[left]  {$1$} (node11);
    \draw [arc] (node1)  -- node[right] {$2.1$} (node12);
    \draw [arc] (node1)  -- node[right] {$3$} (node13);
    \draw [arc] (node11) -- node[left]  {$1$} (node111);
    \draw [arc] (node11) -- node[right] {$2$} (node112);
    \draw [arc] (node12) -- node[right] {$1$} (node121);
    \draw [arc] (node111) -- node[left] {$1$} (node1111);
  \end{tikzpicture}
} \quad
\subfigure[A stripped pid-tree.]
{\label{fig::stripped::e}
  \begin{tikzpicture}[yscale=-0.9,xscale=2]

    \node[treenode,label=right:{}] (root)      at (0.5, 0.5) {};
    \node[treenode,label=right:{}] (node1)     at (0.5, 1.5) {};
    \node[treenode,label=left:{},fill=lightgray] (node13)    at (0, 2.5) {};
    \node[treenode,label=right:{$M_1$}] (node11)    at (0.5, 2.5) {};
    \node[treenode,label=right:{$M_2$}] (node12)    at (1, 2.5) {};
    \node[treenode,label=right:{}, fill=lightgray] (node111)   at (0.5, 3.5) {};
    \node[treenode,label=right:{}, fill=lightgray] (node121)   at (1, 3.5) {};

    \node at (0.5, 4.5) {$M_1 \defeq \{(1.1.1, 1.2.1)\}$};
    \node at (0.5, 5) {$M_2 \defeq \{(1.2.1, 1)\}$};

    \draw [arc] (root)   -- node[right]  {$1$}   (node1);
    \draw [arc] (node1)  -- node {$1.1$} (node11);
    \draw [arc] (node1)  -- node[right] {$2.1$} (node12);
    \draw [arc] (node1)  -- node[left] {$3$}   (node13);
    \draw [arc] (node11) -- node[right]  {$1$}   (node111);
    \draw [arc] (node12) -- node[right] {$1$}   (node121);
  \end{tikzpicture}
} \quad
\subfigure[A stripped pid-tree.]
{\label{fig::stripped::f}
  \begin{tikzpicture}[yscale=-0.9,xscale=2]

    \node[treenode,label=right:{}] (root)      at (0.5, 0.5) {};
    \node[treenode,label=right:{}] (node1)     at (0.5, 1.5) {};
    \node[treenode,label=left:{},fill=lightgray] (node13)    at (0, 2.5) {};
    \node[treenode,label=right:{$M'_2$}] (node11)    at (0.5, 2.5) {};
    \node[treenode,label=right:{$M'_1$}] (node12)    at (1, 2.5) {};
    \node[treenode,label=right:{}, fill=lightgray] (node111)   at (0.5, 3.5) {};
    \node[treenode,label=right:{}, fill=lightgray] (node121)   at (1, 3.5) {};

    \node at (0.5, 4.5) {$M'_2 \defeq \{(1.1.1, 1)\}$};
    \node at (0.5, 5) {$M'_1 \defeq \{(1.2.1, 1.1.1)\}$};

    \draw [arc] (root)   -- node[right]  {$1$}   (node1);
    \draw [arc] (node1)  -- node  {$1.1$} (node11);
    \draw [arc] (node1)  -- node[right] {$2.1$} (node12);
    \draw [arc] (node1)  -- node[left] {$3$}   (node13);
    \draw [arc] (node11) -- node[right]  {$1$}   (node111);
    \draw [arc] (node12) -- node[right] {$1$}   (node121);
  \end{tikzpicture}
}

\caption{Two pid-trees which are not equivalent (\ref{fig::stripped::a},
\ref{fig::stripped::b}) but become equivalent when transformed into
stripped pid-trees (\ref{fig::stripped::c},
\ref{fig::stripped::d}). The pids-trees in Figure
\ref{fig::stripped::e} and \ref{fig::stripped::f} are not equivalent
but denote equivalent states. Most of markings have been omitted, black nodes
represent inactive pids, gray nodes represent next pids.}
\label{fig::stripped}
\end{figure}

This second canonical form will lead to more symmetry detection. This
may seem counterintuitive since we remove data from the tree, but
removing these data also removes constraints we add on pids. These
constraints are introduced by inactive pids that do not appear in the
marking and removing them lead to a better characterisation of the
relations between pids. Because inactive pids do not appear in the original
marking, they should not impact on symmetry detection. Stripping a
pid-tree will remove these pids. An example is given in Figure
\ref{fig::stripped}.

\begin{proposition} \label{prop::uniquestrip}
  Let $M$ be a reachable marking of a t-net $N$. There is a unique
  stripped pid-tree in $\repr(M)$.
\end{proposition}

Even if removing inactive pids allows to provide a better
characterisation of markings, some symmetries may still miss. An
example is given in Figure \ref{fig::stripped::e} and
\ref{fig::stripped::f}, the trees are not equivalent but the
corresponding states are, because pids $1.1$ and $1.2$ are inactive
the relation between them can be released (they cannot be compared
within the model) and thus $1.1.1$ and $1.2.1$ can exchange roles. So
we now introduce a sufficient condition to detect all symmetries with
pid-trees.

\subsection{Sufficient condition for a unique representation}
\label{sec::sufficientcondition}

To start with, we introduce the notion of \emph{clean marking}. A
marking is \emph{clean} if for every active pid, or next-pid, its
parent is active, except for the pid $1$ ``the father of all pids''.
This notion have several consequences. First, there will be a unique
representation of a marking, \ie, $\repr(M) = \{ R(M) \}$. The second
consequence is that we will have:
$$
\pid_{q_M} \cup \nextpid_{q_M}
= \pid(R(M))
= \subpid_{q_M} \cup \nextpid_{q_M}
$$
So the stripped and expanded representations of the marking are equal.

\begin{proposition} \label{prop::cleanunique}
  Let $M$ be a reachable clean marking of a t-net. Then there is a unique
  pid-tree representation of $M$.
\end{proposition}

Now we show that the bijection used to check pid-tree equivalence
is exactly the bijection used to check equivalence between markings.

\begin{proposition} \label{prop::equivimplpath}
  Let $M_1$, $M_2$ be two reachable markings in a t-net. If $M_1, M_2$
  are clean and $M_1 \sim M_2$ then there is a bijection $h$ defined as
  $$ h = \{ (\pi, \pi') ~|~ \pi \in \pid(R(M_1)),~ \pi' \in
  \pid(R(M_2)) ~\mbox{and}~ \relpath{\pi, R(M_1)} = \relpath{\pi',
    R(M_2)}\}
  $$
  such that $M_1 \sim_h M_2$.
\end{proposition}

This proposition states that paths inside the pid-tree representation
perfectly characterise the bijection between states corresponding to
clean markings. It implies that if our pid-trees are equivalent then
the bijection used to check pid-tree equivalence also satisfies the
state equivalence requirements.

\begin{theorem} \label{thm::equiviffequiv}
  Let $M_1$ and $M_2$ be two reachable markings. If $M_1$ and $M_2$
  are clean then $R(M_1) \siblingequiv R(M_2)$ iff $M_1 \sim M_2$.
\end{theorem}

When addressing clean markings we detect all symmetries in state
spaces. It may seem a big constraint to work with clean markings,
however it can be ensured by restriction at modelling level. One can
for example force a process to wait its children to finish before
dying; in these cases every reachable marking will be clean.

Moreover, when addressing arbitrary systems, we can use, for example,
stripped pid-trees to detect potential symmetries with the guarantee
that for every discovered clean marking, we will detect all symmetries.

\section{Conclusion}

We have developed a theoretical package for symmetries detection in
Petri nets with dynamic process creation. This approach is not
complete however it leads to effective algorithms that can be
implemented in a real-life model-checker. We have also shown its
limits and discussed its application and the completeness of symmetry
detection. This has been made using a tree data structure, the
\emph{pid-trees}, and an equivalence relation among them. The
development of this package have been done in a general scope which
results in potential symmetry detection. Then we have looked for
canonisation with \emph{expanded} and \emph{stripped} pid-trees in
order to get a unique representation of a marking. Finally we
expressed a sufficient condition to detect all symmetries within a
model.  Using pid-trees leads to time efficient algorithms to detect
symmetries during state space exploration, which is not the case when
resorting to graph isomorphism computation
\cite{KKPP08,KKPP09,KKPP10}.

The different canonical forms we have introduced allow to implement
the approach and detect symmetries. Even if all symmetries are not
detected we believe that many of them are captured. We will try to
confirm this statement in future works. We intend to implement our
reductions by symmetries within \emph{Neco} net compiler
\cite{FP11a,FP11}. It will lead to case studies and several benchmarks
to validate our intuition about expected performances, however we will
not be able to test every existing model. This is why the sufficient
condition we have introduced in section \ref{sec::sufficientcondition}
is important. It guarantees that we will detect all symmetries in a
class of models, moreover if the condition is not satisfied the method
remains valid even if we lose this guarantee of full detection.

We will also explore possibilities given by the space between our two
bounds, expanded and stripped pid-trees, using more precise labelling
functions to capture symmetries, and address systems without relations
$\sibling_1$ and $\sibling$. For example, pid-trees in Figures
\ref{fig::stripped::e} and \ref{fig::stripped::f} become equivalent
using a labelling function which takes in account the size, or
contents, of markings.

The trade-off between complexity of the computation and completeness
was the main motivation in this work. Indeed developing a complete
approach with an exponential complexity in time would not be usable in
a real life model-checker where the number of discovered states is
almost always exponential in the size of the model. Even if reduction
by symmetries may exponentially reduce the size of the state space,
such an expensive approach will remain not usable due to the number of
states as explained in the introduction. Thus partial methods like
this one should be considered in practice even if they do not capture
every symmetry.

\bibliographystyle{eptcs}
\bibliography{fronc_sub1}

\end{document}